\documentclass[conference]{IEEEtran}
\IEEEoverridecommandlockouts
\usepackage[left=0.625in,right=0.625in,top=0.75in,bottom=1.1in]{geometry} % Adjust the side margins
\usepackage{amsthm}
\usepackage{amssymb}
\usepackage{algorithmic}
\usepackage[inline]{enumitem}
\usepackage{graphicx}
\usepackage{hyperref}%
\usepackage{tikz}
\usetikzlibrary{arrows}
\usepackage{verbatim}
\usepackage{amsmath}
\usepackage[ruled,vlined,linesnumbered]{algorithm2e}
\usepackage{array}
\usepackage{float}
\usepackage{graphicx}
\usepackage{url}
\usepackage{multirow}   
\usepackage{pgfplots}
\usepackage{multicol}
\usepackage{tcolorbox}
\usepackage[export]{adjustbox}
\usepackage{float}
\usetikzlibrary{positioning, fit, calc, shapes, arrows}
\usepackage[underline=false]{pgf-umlsd}

\usepackage{graphicx}
\usepackage{caption, balance}
\usepackage{array,amsfonts}
\usepackage{subcaption}
\usepackage{cite}
\usepackage{wrapfig}
\usepackage{tabularx}
\usepackage[english]{babel}	
\usepackage{soul}
\usepackage{tikz}

\usepackage{epstopdf}
\theoremstyle{remark}

\if CLASSINFOpdf

\else

\fi

\usetikzlibrary{calc}
\usetikzlibrary{shapes.arrows}
\usetikzlibrary{positioning,fit,backgrounds}

\begin{document}

\title{Next-Gen Space-Based Surveillance: Blockchain for Trusted and Efficient Debris Tracking}

\author{\IEEEauthorblockN{Nesrine Benchoubane\IEEEauthorrefmark{1},
Nida Fidan\IEEEauthorrefmark{2}\IEEEauthorrefmark{3},
Gunes Karabulut Kurt\IEEEauthorrefmark{1}, and
Enver~Ozdemir\IEEEauthorrefmark{2}}
\IEEEauthorblockA{\IEEEauthorrefmark{1}Poly-Grames Research Center, Department of Electrical Engineering, Polytechnique Montréal, QC, Canada}
\IEEEauthorblockA{\IEEEauthorrefmark{2} Informatics Institute, Istanbul Technical University, Istanbul, Türkiye}
\IEEEauthorblockA{\IEEEauthorrefmark{3} National Institute of Electronics and Cryptology, TÜBİTAK BİLGEM, Kocaeli, Türkiye}
}

\maketitle

\begin{abstract}
The increasing congestion of Earth's orbit due to growing satellite deployments and space debris poses a significant challenge to sustainable space operations. Traditional space surveillance systems rely on centralized architectures, which introduce single points of failure and scalability constraints. This paper proposes a blockchain-based solution where satellites function as nodes with distinct roles to validate and securely store debris-tracking data. Simulation results indicate that optimal network performance is achieved with approximately 30 nodes, balancing throughput and response time, representing an approximately 9× improvement over traditional consensus mechanisms.\end{abstract}


\IEEEpeerreviewmaketitle

\section{Introduction}

\IEEEPARstart{T}{he} rapid proliferation of satellites, driven by the commercialization and democratization of space, has significantly amplified the threat posed by space debris. By 2023, over 5,000 active satellites orbited the Earth \cite{ucs_satellite_database}, alongside more than 40,500 trackable objects larger than 10 cm \cite{esa_space_debris_2024}, and millions of smaller, untrackable fragments. Traveling at velocities of up to 28,000 km/h, these objects present severe collision risks to both satellites and crewed spacecraft. As the scale and tempo of space operations accelerate, new paradigms for surveillance, coordination, and information sharing are urgently needed to ensure operational safety and long-term sustainability.

\subsection{Space Surveillance and Domain Awareness}
Monitoring and mitigating collision risks in space requires persistent and precise surveillance. This capability is commonly framed under the umbrella of functions of Space Domain Awareness (SDA), which includes the detection, tracking, identification, and prediction of space objects using both ground-based and space-based systems. SDA has traditionally relied on centralized architectures anchored by terrestrial sensors and ground-to-satellite communication. However, this model is increasingly strained by limitations in global coverage, latency, and vulnerability to single points of failure.

Consequently, a shift toward decentralized, space-based architectures is gaining momentum  \cite{gordon2024rolecommunicationsspacedomain}. In this model, satellites are interconnected through inter-satellite links (ISLs) and collaborate to share data and perform in-orbit processing, enabling real-time decision-making and cooperative debris tracking using advanced onboard sensors \cite{9612138, 10663253}. 

\subsection{Data Sharing and Distributed Ledger}
The success of distributed SDA systems hinges on secure, transparent, and efficient data sharing across heterogeneous space assets \textit{in space}. In line with these goals, \cite{10908606, 10535109} proposed an on-orbit marketplace to facilitate data exchange across independently operated systems.

To enable this, we must meet requirements of data integrity, immutability, and transparency—without relying on centralized control—Distributed Ledger Technologies (DLTs) provide a promising solution. These technologies enable decentralized entities to share and verify information in a tamper-resistant manner, while maintaining a consistent system state. Two fundamental capabilities underpin DLTs \cite{9129732, 8416434}: peer-to-peer (P2P) communication, which governs how entities communicate without a central authority, and consensus mechanisms, which allow entities to agree on the accepted version of truth. 

DLT architectures can vary significantly in structure, each with different implications summarized in Table~\ref{tab:dlt}, and fall into three broad categories:

\begin{itemize}
    \item \textbf{Sequential Ledger}: This structure consists of a linear chain of blocks, each containing a set of transactions linked cryptographically to the previous block, which requires formal validation and global consensus.
    \item \textbf{Web Ledger}: This structure forms a directed acyclic graph where each transaction validates previous ones  and thus supports parallel validation and eliminates the need for mining.
    \item \textbf{Centric Ledger}: In this structure, each maintains its own chain of transactions and updates it independently and where synchronization with others occurs asynchronously.
\end{itemize}

\subsection{Debris Tracking}

\begin{table*}[btp!]
\caption{Comparative overview of distributed ledger structure.}
\label{tab:dlt}
\centering
\renewcommand{\arraystretch}{1.3} 
{ 
\begin{tabular}{|l|p{1.2cm}|c|c|p{3.2cm}|p{3.2cm}|}
\hline
 \textbf{Type} & \textbf{Structure} & \textbf{Consensus} & \textbf{P2P} & \textbf{Strength} & \textbf{Limitation} \\
\hline \hline
Sequential Ledger & Linear chain of blocks & Global consensus  & Broadcast & High integrity, auditability, strong trust model  & Limited scalability and energy intensive \\ \hline
Web Ledger& Directed graph & Tip selection & Local propagation (Gossip) & Scalable, no mining required, parallel processing & Depends on activity, orphaned transactions \\ \hline
Centric Ledger & Individual chains & Representative voting & Selective or on-demand sync & Energy-efficient, high autonomy, fast transactions & Complex coordination \\ \hline
\end{tabular}}
\end{table*}

As analyzed from the characteristics of DLT structures, each ledger model maps to specific debris tracking architectures. A sequential ledger is ideal for globally coordinated debris tracking, where a small number of cooperative satellites operated by multiple stakeholders must maintain a tamper-proof, chronologically ordered log. This structure supports formal validation, traceability, and cross-agency trust. A web-based ledger, by contrast, suits large single-operator constellations, where satellites can locally validate observations in a scalable manner without global consensus, enabling rapid, low-latency updates in dense environments. Lastly, a centric ledger allows individual actors to append local records and sync intermittently, making it appropriate for hybrid, multi-orbit systems or fragmented ownership scenarios where asynchronous data reconciliation across orbits is necessary.

\subsection{Motivations and Contributions}

Given our goal of enabling a globally maintained, immutable debris log among trusted stakeholders, we adopt the sequential ledger structure. It aligns with scenarios where satellites are fewer, and cooperation is required.

Prior studies have demonstrated the potential of integrating blockchain in satellite networks. \cite{9520348} introduces communication protocols that leverage blockchain for authentication and privacy in different satellite interaction scenarios. A separate approach \cite{10219188} explores capability-based access control and automated smart contract security in blockchain-based space systems. Other research \cite{9837853} proposes a DAG-structured blockchain consensus mechanism to improve fault tolerance and reputation assessment in satellite networks through distributed computing. Additional conceptual frameworks, such as BESTA, focus on leveraging blockchain for behavioral anomaly detection in space traffic \cite{REED}.

Existing work mainly focuses on general blockchain integration and does not address the specific requirements of space surveillance systems for debris detection. These studies often overlook how traditional consensus mechanisms can be optimized for such networks without interfering with their core functions. Our work fills this gap by tailoring blockchain consensus mechanisms to ensure optimal performance and thus, we make the following contributions:
\begin{itemize}
    \item We propose a network model that integrates with the current state-of-the-art optical decentralized surveillance network;
    \item We leverage blockchain technology to ensure secure and tamper-resistant record-keeping to maintain data integrity and traceability;
    \item We develop a role-based consensus mechanism that optimizes network performance by assigning dedicated roles for verification and approval phases.
\end{itemize}

% \subsection{Structure}
% The rest of this paper is organized as follows. Section \ref{Preliminary} presents preliminary concepts. Section \ref{use-case} presents the system model. Section \ref{Method} presents the mathematical model. Section \ref{Results} presents the numerical results. Section \ref{big-pic} discusses the resulting integrated network. Section \ref{Conclusion} concludes the paper.

% \hl{cast to related work}

\section{Preliminary Concepts}
\label{Preliminary}

\tikzset{
  circlelabel/.style={
    fill=gray!90, 
    text=black, 
    font=\tiny\bfseries, 
    inner sep=1mm, 
    minimum size=1mm, 
    shape=circle
  }
}

\newcommand*\circled[1]{\tikz[baseline=(char.base)]{
    \node[circlelabel] (char) {#1};}}
\begin{figure}[tb]
\centering
\begin{tikzpicture}[
    node distance=2.5cm, 
    every node/.style={draw, rounded corners, text centered, minimum height=1cm, draw=none, font=\footnotesize},
    arrow/.style={thick,->,>=stealth},
    greybox/.style={fill=gray!20, text width=9cm, align=justify, inner sep=5mm, font=\normalsize},
    circlelabel/.style={fill=gray!90, text=black, font=\tiny\bfseries, inner sep=1mm, minimum size=1mm, shape=circle} % Circle style
]

% Nodes
\node (a) {\includegraphics[width=0.7cm]{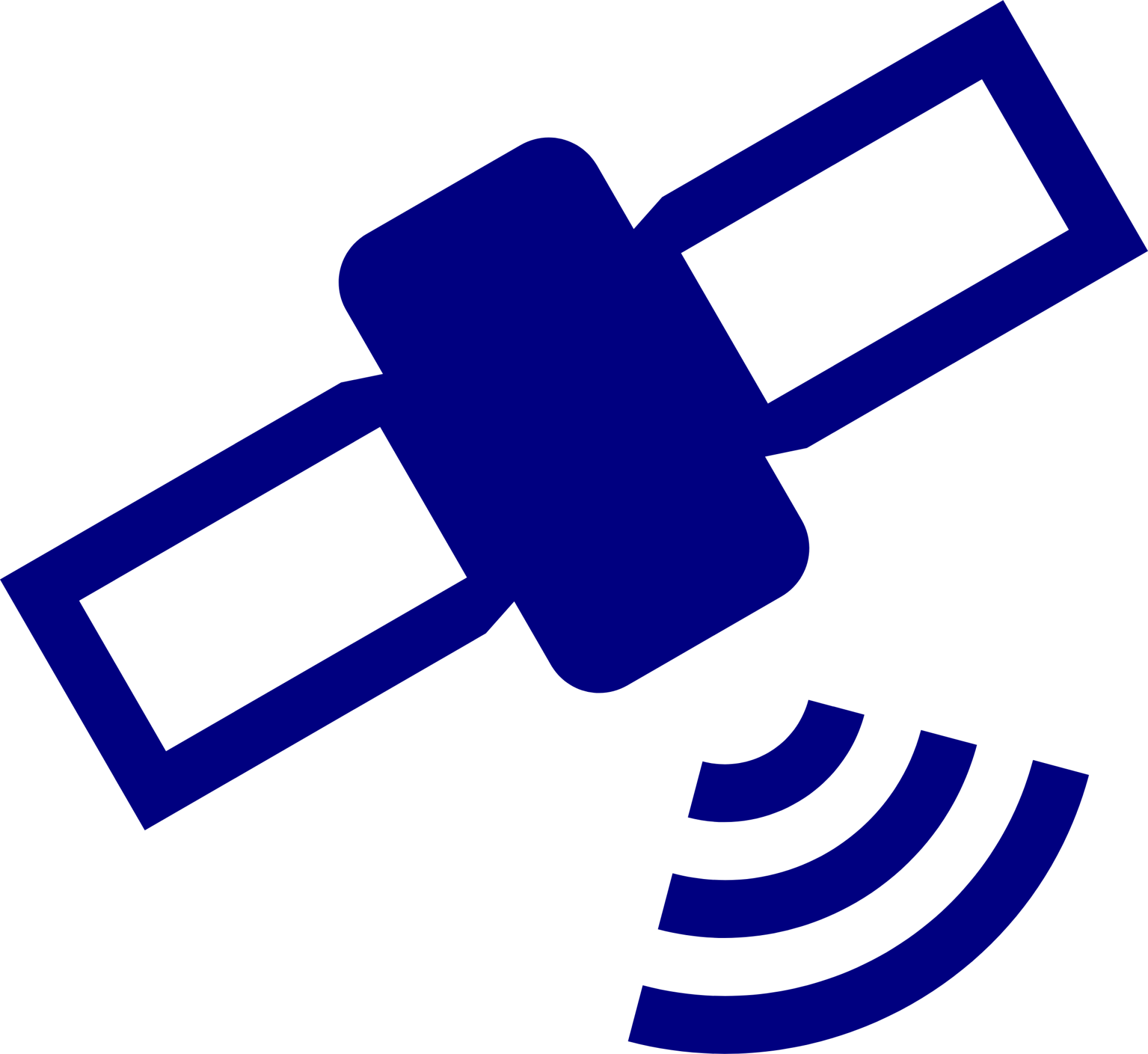}};
\node[below=0.2mm of a] {Node};
\node (b) [right of=a] {\includegraphics[width=1.2cm]{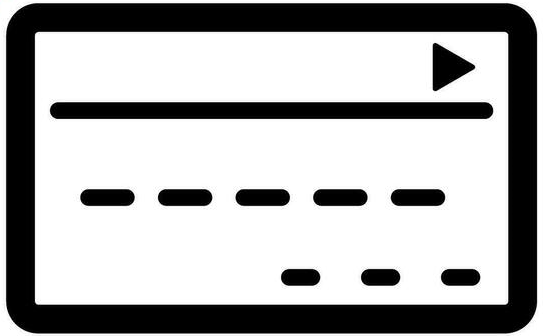}};
\node[below=0.1mm of b] {Transaction};
\node (c) [right of=b] {\includegraphics[width=1.4cm]{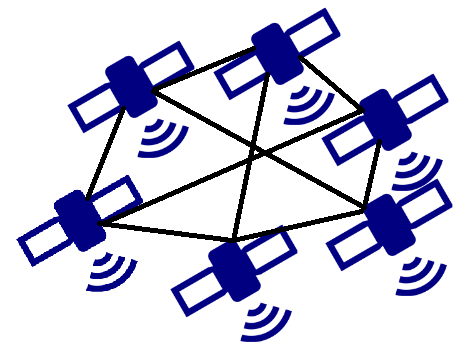}};
\node (d) [right of=c] {\raisebox{4mm}{\includegraphics[width=1.5cm]{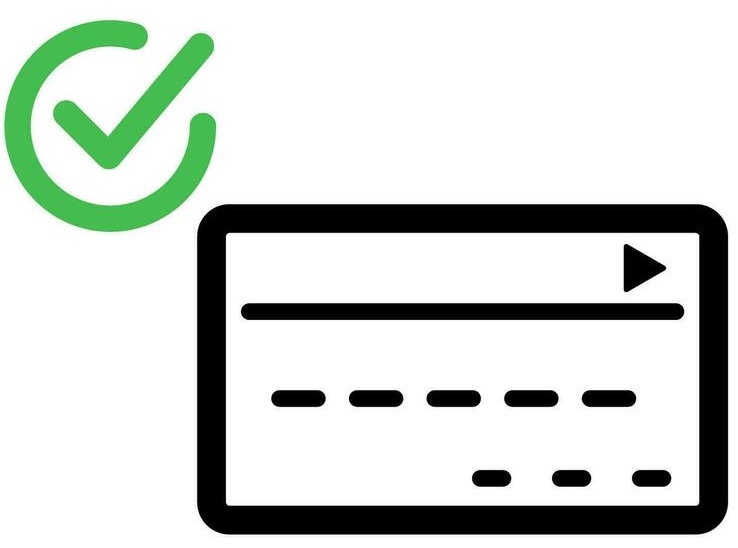}}};
% \node[right=0.2mm of d] {Verification phase};
\node (e) [below of=d] {\includegraphics[width=1.5cm]{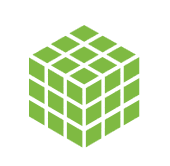}};
\node[below=0.1mm of e] {New block};
\node (f) [left of=e] {\includegraphics[width=1.5cm]{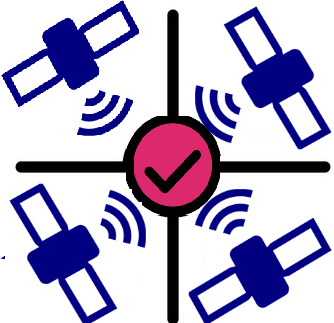}};
% \node[right=0.2mm of f] {Consensus phase};
\node (g) [below of=f] {\includegraphics[width=3.5cm]{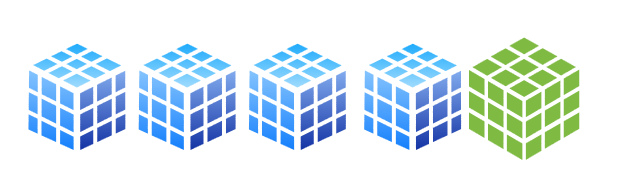}};

\draw[arrow] (a) -- node[left, xshift=-2.1mm, circlelabel] {1} (b);
\draw[arrow] (b) -- node[left, xshift=-1mm, circlelabel] {2} (c);
\draw[arrow] (c) -- node[left, circlelabel] {3} (d);
\draw[arrow] (d) -- node[above, circlelabel] {4} (e);
\draw[arrow] (e) -- node[right, circlelabel] {5} (f);
\draw[arrow] (f) -- node[above, circlelabel] {6} (g);

\end{tikzpicture}
\begin{tikzpicture}[
    greybox/.style={draw=none, fill=white, align=left, minimum width=4.5cm, text height=1.5ex, text depth=0.5ex, inner sep=2pt, font=\small},
    circlelabel/.style={fill=gray!90, text=black, font=\tiny\bfseries, inner sep=2pt, shape=circle},
    bigbox/.style={draw=none, fill=gray!50, inner sep=4pt},
    node distance=2mm and 2mm
]

\node[circlelabel] (step1) {1};
\node[greybox, right=of step1, anchor=west] (desc1) { A node initiates a transaction by creating a data record.};

% Step 2
\node[circlelabel, below=of step1] (step2) {2};
\node[greybox, right=of step2, anchor=west] (desc2) { The transaction is broadcast across the network.};

% Step 3
\node[circlelabel, below=of step2] (step3) {3};
\node[greybox, right=of step3, anchor=west] (desc3) { Nodes verify the transaction to ensure it is valid.};

% Step 4
\node[circlelabel, below=of step3] (step4) {4};
\node[greybox, right=of step4, anchor=west] (desc4) { Verified transactions are grouped into a new block.};

% Step 5
\node[circlelabel, below=of step4] (step5) {5};
\node[greybox, right=of step5, anchor=west] (desc5) {The block undergoes a consensus process for approval.};

% Step 6
\node[circlelabel, below=of step5] (step6) {6};
\node[greybox, right=of step6, anchor=west] (desc6) {Once consensus is reached, the block is added to the chain.};
\end{tikzpicture}
\caption{Blockchain functional flow.
}
\label{fig:blockchain functional flow}
\end{figure}

As part of the sequential ledger mechanism, the blockchain functional flow is illustrated in Fig.~\ref{fig:blockchain functional flow}. First, the nodes initiate and submit transactions to the network (\circled{1} which takes milliseconds to one second). A transaction represents a record of an action or exchange, such as transferring data or assets, which needs to be validated before being permanently recorded (\circled{2} which takes milliseconds to a few seconds). Submitted transactions are broadcast across the network and temporarily stored in a pool of pending transactions (\circled{3} which takes milliseconds to a few seconds). These transactions are grouped into blocks by designated nodes (\circled{4} which takes milliseconds to a few seconds). Before a block can be added to the blockchain, it undergoes an approval process through a consensus algorithm (\circled{5} which takes seconds to minutes). This mechanism ensures that all participating nodes agree on the validity of the block and its transactions. Only after this consensus is reached does the block become a permanent part of the blockchain (\circled{6} which takes milliseconds to a few seconds).

In our interest are the consensus algorithms that govern the approval of each block (\circled{5}). While consensus ensures agreement across distributed nodes, traditional algorithms present limitations in performance and scalability under constrained environments. Recent studies, including \cite{JAIN2025100065}, provide throughput benchmarks across various protocols. PBFT, for instance, achieves over 1000 transactions per second (TX/s) with fewer than 20 nodes but suffers from communication overhead. Other protocols such as DPoS, PoET, Raft, and ByzCoin require over 1000 nodes to match this throughput, making them less feasible for systems with tight resource and connectivity constraints. This motivates the development of consensus mechanisms tailored for reduced node participation, intermittent links, and latency-sensitive applications, as targeted in our design.

\section{Network Model}
\label{use-case}

\begin{figure}[t!]
\centering
\includegraphics[width=\linewidth]{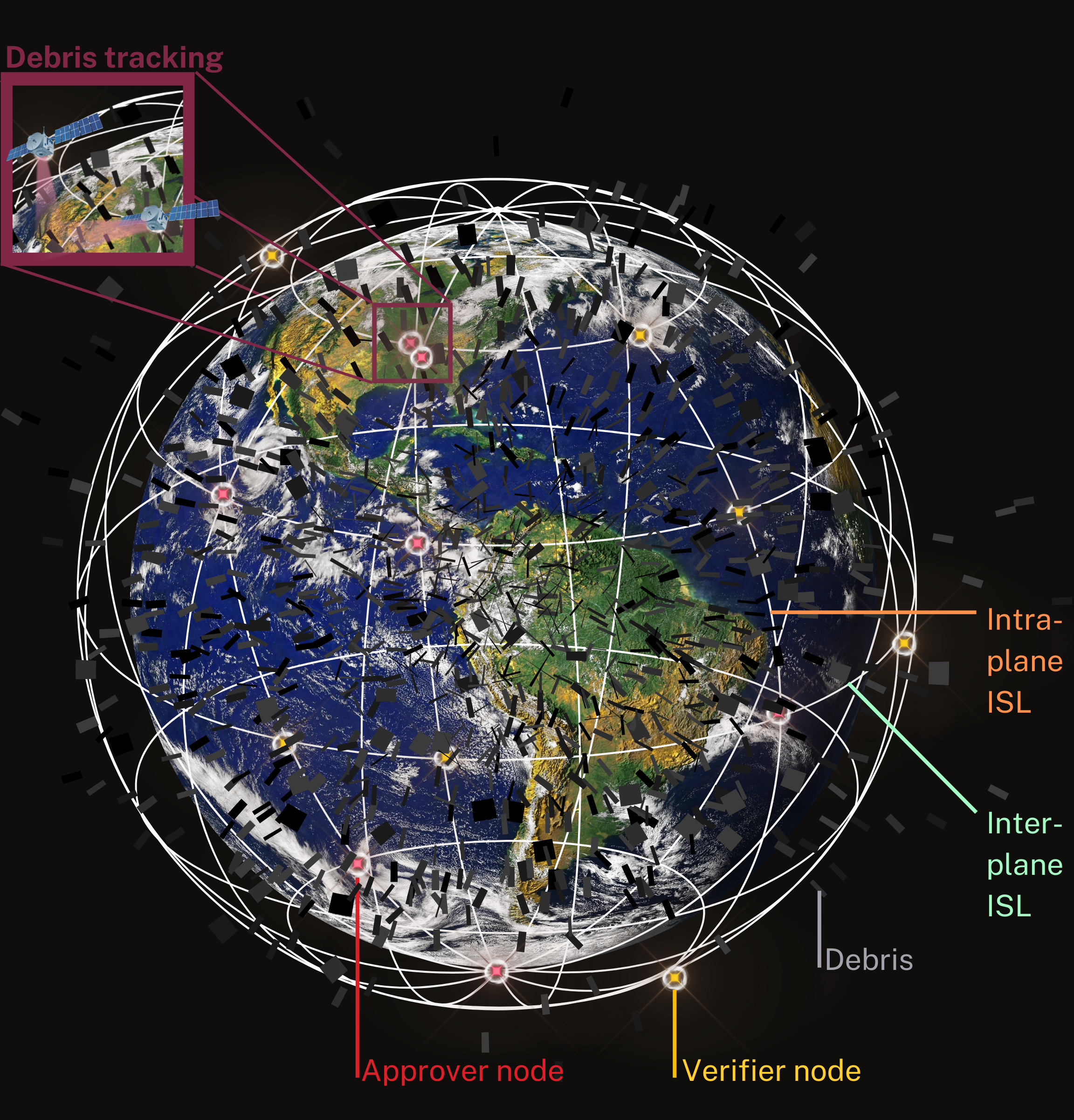}
\caption{Overview of the proposed network.}
\label{fig:nodesnetwork}
\end{figure}

Recent advancements in optical space-based surveillance networks have demonstrated significant improvements in tracking LEO debris. In \cite{10663253}, a multi-satellite optical tracking network was proposed, operating with a measurement interval of 120 seconds, and utilizing multiple point-of-view (MPOV) observations for more accurate orbit determination. The optimal network consists of 24 satellites distributed, ensuring continuous coverage of the 700–1,000 km altitude range. The study also found that a 120-second interval provided the lowest mean errors. Building upon this foundation, we propose a secure and efficient blockchain space-based network aimed at enhancing the integrity and reliability of debris tracking data, as illustrated in Fig. \ref{fig:nodesnetwork}.

\subsection{Key Constraints for Validation and Design}
We must meet the two following constraints, as expressed in Eq. \ref{constraints}, to ensure our network aligns with the design and operates effectively within the established optimal points of the optical network:
\begin{enumerate}[label=(\roman*)]
    \item \textbf{Satellite Count}: The baseline network consists of 24 satellites. If the number of satellites in our network exceeds this threshold, we must apply network partitioning strategies to maintain performance.
    \item \textbf{Total Delay}: The total delay incurred by all operations performed in our algorithm must be completed within the 120-second measurement window. This ensures that the blockchain operations do not interfere with the core data collection process and maintain synchronization with the optical surveillance network.
\end{enumerate}

\begin{equation}
\left\{
    \begin{array}{ll}
         (i)    & n <= N = 24  \\
         (ii)     & t_{\text{blockchain}} \leq T = 120 ~\textrm{s} 
    \end{array}
\right.
\label{constraints}    
\end{equation}

\subsection{Roles}
Satellites (also referred to as nodes) within the network perform different roles to facilitate consensus. A node has to take at least one of the following roles:

\paragraph{Approver} responsible for validating blocks and deciding whether a new block can be added to the blockchain. Only the approvers participate in the approval process of a block.

\paragraph{Verifier} responsible for authenticating network packets to ensure the integrity and correctness of the data transmitted across the network.

\subsection{Links}
Satellites within the network communicate using ISLs which serve as the backbone for data exchange, consensus synchronization, and distributed verification tasks. These links are categorized as:

\subsubsection{Intra-plane ISLs} These are permanent communication links established between satellites within the same orbital plane, where the satellites maintain continuous visibility of one another. 

\subsubsection{Inter-plane ISLs} These links are established between satellites in different orbital planes, but are typically temporary and subject to periodic interruptions. Due to the dynamic nature of orbital motion, these links cannot be maintained permanently, and their availability depends on the relative positioning of the satellites.

\subsection{Configuration}

In this work, we investigate two satellite network configurations: Walker Delta and Walker Star. These configurations differ in their satellite distribution schemes and thus also impact the communication latency. Walker Delta places satellites evenly across several orbital planes with specific phase shifts, typically optimizing coverage at mid-latitudes. In contrast, Walker Star employs a more symmetric distribution of orbital planes, enabling more uniform global coverage, including improved service in polar regions.

\section{Consensus Model} 
\label{Method}

The proposed consensus model operates in three key phases, where all nodes in the network are assumed to have a trust relationship with each other.

\subsection{Key Generation Phase}

\begin{algorithm}[thp!]
    {A group manager selects a base field $\mathbb F_p$ and a polynomial $f(x)$ of degree $t-1$ over $\mathbb F_p$. }\\
{The group manager determines an elliptic curve $E(\mathbb F_p)$ with a prime order. }\\
	{Each node $U_i$  receives a public identity number, $x_i \in \mathbb F_p$ from the manager. }\\
	{A node $U_i$ also receives its private key $f(x_i)$ from the group manager.} \\
	
	\caption{Key Generation Phase}
    \label{KeyGen}
\end{algorithm}

Let $S=\{U_1, \dots , U_n\}$ denote the nodes in the network. In the key generation phase, described in Algorithm~\ref{KeyGen}, the group manager is responsible for setting up the cryptographic parameters for the network and all its nodes. Specifically:

\begin{itemize}
    \item  \textbf{Public values:} $x_1, \dots , x_n $ are public values (identifying information of nodes);
    \item \textbf{Private values:} $f(x_1), \dots , f(x_n)$ are private values assigned to nodes $U_1, \dots , U_n$ respectively;
    \item \textbf{Elliptic Curve Setup:} The finite field $\mathbb F_p$, where $p$ is a prime number and an elliptic curve $E$ over $\mathbb F_p$. $f(x)$ is a degree $t-1$ polynomial over $\mathbb F_p$ and $x_1,\dots ,x_n$ are elements of a finite field $\mathbb F_p$;
    \item \textbf{Reference Polynomial:} The polynomial $f(x)$ of degree $t-1$. 
\end{itemize}

\subsection{Mining Phase}

\begin{algorithm}[ht!]
	{At least $t$ out of $n$ available nodes join the mining process. They are called approvers.}\\
	{Each miner gets the public identity information of others.}\\
	{Each miner computes:
		$$c_k=\prod_{m=0,m\ne k}^{t}\left(\frac{-x_m}{x_k-x_m}\right)f(x_k)P$$
	for $i=1.\dots,t$ and their public keys are $x_1,\dots,x_r$.}\\
	{Each miner releases their $c_k$ value publicly to be included in the block.}
	\caption{Mining Phase}
    \label{Mining Phase}
\end{algorithm}

Once the keys are established, the approver nodes perform the task of validating and approving blocks. To add a new block to the blockchain, at least $t$ out of $n$ nodes must participate in this phase, detailed in Algorithm~\ref{Mining Phase}. Each approver gets the public identifier of the others. Each computes its $c_k$ values and releases $c_k$ publicly to be included in the block where it can be expressed. When nodes achieve this result, the identity information of the nodes participating in the approval process is included in the block, and added to the blockchain.

\subsection{Verifying Phase}

After the mining phase, the verifier nodes take over in the verifying phase, responsible for ensuring the integrity of the transmitted network packets. Each verifier node ensures that the data being transmitted between nodes is accurate and unaltered. The phase is described in Algorithm~\ref{Verifying Phase} and requires at least $v$ verifier nodes to authenticate the network packets. The verifiers (at least $\deg(f(x))+1$ verifiers needed) compute the sum of $c_k$'s and confirm the result is $f(0)P$. Then they sign the results by computing $d_k$ and get the sum of all and put the sum in the block. 

\begin{algorithm}[ht!]
    {Suppose that $v$ nodes perform the verification: Assume their public keys: $y_1,\dots,y_v$ where $v\ge t$. They are called verifier nodes.} \\
	{Each verifier node computes the sum of $c_i$'s and confirms the result is $f(0)P$.}\\
	{In order to sign the result, each verifier node computes $$d_k=\prod_{m=0,m\ne k}^{v}\left(\frac{-y_m}{y_k-y_m}\right)f(y_k)P$$}\\
	{Each verifier node releases its result $d_k$ to the public.}

	\If{${\overset{r}{\underset{i=1}{{\displaystyle\sum}}}c_i}$ is equal to ${\overset{v}{\underset{i=1}{{\displaystyle\sum}}}d_i}$}
	{
		  "Verification is valid!"
	}
	\Else{
		"Repeat Verification Phase!"
	}

	\caption{Verifying Phase}
    \label{Verifying Phase}
\end{algorithm}

\section{Numerical Results}
\label{Results}

The experiments were conducted on a Windows 10 machine with an AMD Ryzen 7 5800HS processor (3.20 GHz) and 16 GB of RAM. First, we examine the performance of two distinct satellite network configurations with Table~\ref{laten} summarizing the latency results for both configurations across three different orbital planes. 

\begin{table}[tbp!]
\centering
\caption{Latency results by constellation configuration for 3 orbital planes.}
\label{laten}
\renewcommand{\arraystretch}{1.3} 
\begin{tabular}{|l|p{1cm}|p{1cm}|p{1.5cm}|p{1.5cm}|p{1.5cm}|}
\hline
 \textbf{Configuration} & \textbf{Sats per Orbit} & \textbf{Total Sats} & \textbf{Intra-ISL Latency (ms)} & \textbf{Inter-ISL Latency (ms)} \\ \hline\hline
\textbf{Walker Delta}   & 8                        & 24                  & 17.67                        & 17.06                        \\ 
                        & 11                       & 33                  & 13.01                        & 14.77                        \\ 
                        & 13                       & 39                  & 11.05                        & 14.15                        \\ 
                        & 15                       & 45                  & 9.59                        & 14.12                        \\ \hline
\textbf{Walker Star}     & 8                        & 24                  & 17.67                        & 8.59                         \\ 
                        & 11                       & 33                  & 13.01                        & 14.76                        \\ 
                        & 13                       & 39                  & 11.05                       & 9.82                         \\ 
                        & 15                       & 45                  & 9.60                         & 9.09                         \\ \hline
\end{tabular}
\end{table}

% \cite{Kur2021} isls

In both configurations, adding more satellites results in decreased intra-ISL latency, as more links between satellites are established within each orbital plane. In the Walker Delta configuration, while intra-ISL latency decreases with more satellites, the inter-ISL latency remains relatively constant. This indicates that communication between satellites in different planes still experiences similar delays, regardless of the total number of satellites. However, in the Walker Star configuration, both intra-ISL and inter-ISL latencies decrease as more satellites are added. This reduction in inter-ISL latency is particularly significant, as it improves the efficiency of communication between planes, which is crucial for the consensus process.

In the next phase, we implement the proposed consensus model, written in Go and executed within the Windows Subsystem for Linux (WSL). The simulation parameters used in these experiments are summarized in Table~\ref{table:simulation-parameters}. 

\begin{table}[ht!]
\caption{Simulation parameters.}
\label{table:simulation-parameters}
\centering
\renewcommand{\arraystretch}{1.5} 
\begin{tabular}{|p{2cm}|p{3.5cm}|p{2cm}|}
\hline
 \textbf{Parameter}  & \textbf{Description} & \textbf{Value} \\ \hline \hline
 Size of batch   & Specifies the number of transactions or messages processed together in a batch.   & 500            \\ \hline
 Size of message & Defines the size of each message buffer.                                      & 1024 bytes     \\ \hline
 Number of cores    & Represents the number of CPU cores being utilized for the workload.  & Scales with network size \\ \hline
\end{tabular}
\end{table}

To evaluate the network's performance, we use two key metrics: throughput and response time. We then test our consensus model, which incorporates varying role distributions, against state-of-the-art approaches that rely on full participation. Different configurations were tested by adjusting the percentage of verifiers, approvers, and shared-role nodes. The role assignment strategies are outlined in Table \ref{tab:nodes}.

\begin{table}[ht!]
\caption{Roles distribution for varying total node count.}
\label{tab:nodes}
\centering
\renewcommand{\arraystretch}{1.5} 
{ 
\begin{tabular}{|c|c|c|c|}
\hline
 \textbf{Total} & \textbf{Verifiers} & \textbf{Approvers (30\%)} & \textbf{Shared Roles (20\%)} \\
\hline \hline
15  & 14  & 4  & 3  \\
 
20  & 18 & 6  & 4  \\
25  & 22 & 8  & 5  \\
 
30  & 27 & 9  & 6  \\
35  & 31 & 11 & 7  \\
 
40  & 36 & 12 & 8  \\
45  & 41 & 13 & 9  \\
 
50  & 45 & 15 & 10 \\
\hline
\end{tabular}
}
\end{table}

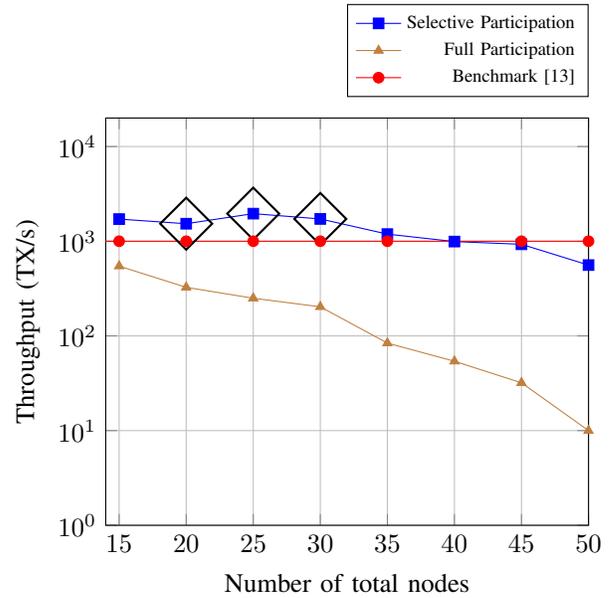
\begin{figure}[tb!]
\centering
    \begin{tikzpicture}
    \begin{axis}[
        ylabel={Throughput (TX/s)},
        xlabel={Number of total nodes},
        ylabel near ticks,
        ymode=log,
        log basis y=10,
        ymin=1, ymax=20000,
        xmin=14, xmax=50,
        ytick={1, 10, 100, 1000, 10000},
        grid=both,
        height=6.5cm,
        width=8cm,
        legend style={at={(0.75, 1.28)}, anchor=north, 
        legend cell align=right, font=\scriptsize}, % Position and left-align legend
    ]

    \addplot[color=blue, mark=square*] coordinates {
        (15, 1713) (20, 1531) (25, 1956) (30, 1720) (35, 1189) (40, 992)  (45, 928) (50, 560)
    };

    \addplot[color=brown, mark=triangle*] coordinates {
        (15, 546) (20, 326) (25, 250) (30, 203) (35, 84) (40, 54)  (45, 32) (50, 10)
    };
    \addplot[color=red,mark=*] coordinates {
        (0, 1000)  (15, 1000) (20, 1000) (25, 1000) (30, 1000) (35, 1000) (1000)  (45, 1000) (50, 1000)
    };
    \addlegendentry{Selective Participation}
    \addlegendentry{Full Participation}
    \addlegendentry{Benchmark \cite{JAIN2025100065} }
      \node (mark) [draw, black, diamond, minimum size = 5pt, inner sep=5pt, thick] 
      at (axis cs: 25, 1956) {};
      \node (mark) [draw, black, diamond, minimum size = 5pt, inner sep=5pt, thick] 
      at (axis cs: 30, 1720) {};
      \node (mark) [draw, black, diamond, minimum size = 5pt, inner sep=5pt, thick] 
      at (axis cs: 20, 1531) {};
    \end{axis}
\end{tikzpicture}
\caption{Throughput results for different network sizes.}
\label{fig:throughput-results}
\end{figure}

\begin{figure}[tb!]
\centering
    \begin{tikzpicture}
    \begin{axis}[
        ylabel={Response time (s)},
        xlabel={Number of total nodes},
        ylabel near ticks,
        ymode=log,
        log basis y=10,
        ymin=1, ymax=150,
        xmin=14, xmax=50,
        grid=both,
        height=6.5cm,
        width=8cm,
        legend style={at={(0.75, 1.2)}, anchor=north, 
        legend cell align=right, font=\scriptsize}, % Position and left-align legend
    ]

    \addplot[color=blue, mark=square*] coordinates {
       (15, 5.39)  (20, 6.20) (25, 6.87) (30, 5.37) (35, 5.38) (40, 6.69)  (45, 5.61) (50, 6.45)
    };
    \addplot[color=brown, mark=triangle*] coordinates {
       (15, 6.85)  (20, 25.57) (25, 37.9) (30, 39.13) (35, 40.65) (40, 48.67)  (45, 49.2) (50, 58.3)
    };
    \addlegendentry{Selective Participation}
    \addlegendentry{Full Participation}

      \node (mark) [draw, black, diamond, minimum size = 5pt, inner sep=5pt, thick] 
      at (axis cs: 25, 6.87) {};
      \node (mark) [draw, black, diamond, minimum size = 5pt, inner sep=5pt, thick] 
      at (axis cs: 30, 5.37) {};
      \node (mark) [draw, black, diamond, minimum size = 5pt, inner sep=5pt, thick] 
      at (axis cs: 20, 6.20) {};
    \end{axis}
\end{tikzpicture}
\caption{Response time results for different network sizes.}
\label{fig:response time-results}
\end{figure}
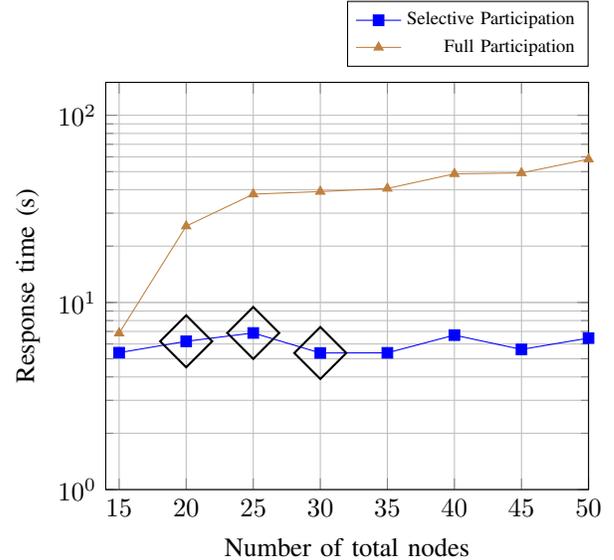
Throughput results, depicted in Fig. \ref{fig:throughput-results}, serve as the primary metric for assessing the network's capacity to handle transactions. It is calculated by dividing the total number of transactions successfully committed by the execution time measured in seconds. Using the results in \cite{JAIN2025100065} as a benchmark, our configurations achieve a notable throughput of 1531 TX/s with fewer than 20 nodes, outperforming PBFT, which is reported to reach 1000 TX/s. Notably, no benchmark results are available beyond 20 nodes; however, we have extended the benchmark up to 50 nodes.

Additionally, the results for the selective participation configuration show that throughput increases from 1391 TX/s at 15 nodes to 1931 TX/s at 30 nodes. Beyond this, throughput declines, dropping to 560 TX/s at 50 nodes. This informs us that, while the network can efficiently process transactions with a moderate number of nodes, scaling beyond this point leads to diminishing returns due to increased communication overhead. Comparatively, in the fully participatory consensus configuration, throughput starts at just 546 TX/s for 15 nodes and decreases rapidly, reaching only 10 TX/s at 50 nodes, thus never meeting the benchmark in \cite{JAIN2025100065}.

Response time is another metric for evaluating the network's efficiency and measures the total delay in this process, encompassing both the verification and approval phases. Results are shown in Fig. \ref{fig:response time-results}. As the number of nodes increases, the response time initially increases for the selective participation configuration, peaking at 6.87 seconds at 25 nodes. The peaks are explained by the higher contention for roles and the relatively smaller pool of nodes handling both the verification and approval processes. 
However, as we increase nodes, the response time drops to 5.37 seconds (30 nodes). In contrast, the configuration of fully participatory consensus sees a sharp increase in response time as the number of nodes increases and then continues to rise, with 58.3 seconds (50 nodes).

In the proposed consensus algorithm, not all nodes participate in the verification and approval processes. Optimal performance is achieved with approximately 20 - 30 nodes (indicated by the symbol $\diamond$ in Fig. \ref{fig:throughput-results} and Fig. \ref{fig:response time-results}), striking a balance between maximum throughput and minimal response time. To satisfy the constraints in Eq. \ref{constraints}, with the optimal points achieved, we have to specify a network of $n = 24 \leq N $ nodes and a response time $t_{\text{blockchain}} = 6.87 ~ s \leq T ~ $. In here, we found that the Walker Star configuration demonstrates the best performance, maintaining low latency across both intra-ISL and inter-ISL communications, consistently under 20 ms.

\section{Big Picture}
\label{big-pic}

We can then describe the three key phases that underpin the operation of the network (integrated with \cite{10663253}) in Fig. \ref{fig:blockchain functional flow}:

\begin{enumerate}[label=\textbf{(Phase \arabic*)}, left=0pt, labelsep=0.4em]
\item \textbf{Data Generation (Debris Tracking)}

Every 120 seconds, satellites collect optical tracking data on debris, which includes position, velocity, and uncertainties. This data is crucial for accurate orbital predictions. It is then packaged as a blockchain transaction to be processed by the network (\circled{1}\(\rightarrow \)\circled{2}).

\item \textbf{Blockchain Processing} 

The 24-node blockchain network processes and stores the tracking data. In this process, the approver satellites perform consensus and add blocks to the blockchain, while the verifiers verify the blocks. This network achieves a throughput of 1956 TX/s within 6.87 seconds (\circled{3} \(\rightarrow \) \circled{5}).

\item \textbf{Real-Time Decision Making}

New debris data is validated and stored in secure blocks under 6.87 seconds, providing near real-time updates well before the next 120-second measurement cycle.
\end{enumerate}

We also highlight several directions for future work based on the assumptions we made and observations gathered:

\begin{itemize}
    \item \textbf{Quantification of Measurement Errors:} The propagation of measurement errors depends heavily on the onboard sensor types, their resolution, calibration, and associated signal processing chains. Future research should consider quantify acceptable error margins and their impact on decision-making and consensus accuracy.
    \item \textbf{Computational and Energy Efficiency:}  The computational load introduced by consensus operations must be balanced against the limited processing capacity and energy budgets of satellites. Evaluating this trade-off requires a joint analysis of onboard processing constraints, solar energy harvesting profiles, and energy storage limits. Future directions include exploring dynamic task offloading to higher-capacity nodes when required.
\end{itemize}

\section{Conclusion}
\label{Conclusion}

This paper presents a novel blockchain-enabled architecture for efficient decentralized space surveillance. Our simulation results indicate that a network under 30 nodes achieves optimal throughput and response time. We also compare our architecture with a fully participatory consensus model, where all nodes perform both verification and approval tasks. Across all scenarios, our approach demonstrates a 9× improvement in both throughput and response time compared to the full participatory consensus, highlighting the efficiency gains achieved by assigning dedicated roles for verification and approval. 

\bibliographystyle{IEEEtran} 
\bibliography{references}

\begin{thebibliography}{10}
\providecommand{\url}[1]{#1}
\csname url@samestyle\endcsname
\providecommand{\newblock}{\relax}
\providecommand{\bibinfo}[2]{#2}
\providecommand{\BIBentrySTDinterwordspacing}{\spaceskip=0pt\relax}
\providecommand{\BIBentryALTinterwordstretchfactor}{4}
\providecommand{\BIBentryALTinterwordspacing}{\spaceskip=\fontdimen2\font plus
\BIBentryALTinterwordstretchfactor\fontdimen3\font minus \fontdimen4\font\relax}
\providecommand{\BIBforeignlanguage}[2]{{%
\expandafter\ifx\csname l@#1\endcsname\relax
\typeout{** WARNING: IEEEtran.bst: No hyphenation pattern has been}%
\typeout{** loaded for the language `#1'. Using the pattern for}%
\typeout{** the default language instead.}%
\else
\language=\csname l@#1\endcsname
\fi
#2}}
\providecommand{\BIBdecl}{\relax}
\BIBdecl

\bibitem{ucs_satellite_database}
\BIBentryALTinterwordspacing
{{Union of Concerned Scientists}}. (2024) \text{UCS Satellite Database}. Accessed: 16 November 2024. [Online]. Available: \url{https://www.ucsusa.org/resources/satellite-database}
\BIBentrySTDinterwordspacing

\bibitem{esa_space_debris_2024}
\BIBentryALTinterwordspacing
{{European Space Agency}}. (2024) \textit{Space debris by the numbers}. Accessed: 16 November 2024. [Online]. Available: \url{https://www.esa.int/Space\_Safety/Space\_Debris/Space\_debris\_by\_the\_numbers}
\BIBentrySTDinterwordspacing

\bibitem{gordon2024rolecommunicationsspacedomain}
N.~G. Gordon \emph{et~al.}, ``{On the Role of Communications for Space Domain Awareness},'' 2024, arXiv: 2406.05582.

\bibitem{9612138}
H.~Yunpeng, L.~Kebo, L.~Yan'gang, and C.~Lei, ``{Review on strategies of space-based optical space situational awareness},'' \emph{Journal of Systems Engineering and Electronics}, vol.~32, no.~5, pp. 1152--1166, 2021.

\bibitem{10663253}
S.~Hamidian, A.~R. Kosari, and N.~Assadian, ``{An Optical Space-Based Surveillance Network for Tracking {LEO} Debris},'' \emph{IEEE Aerospace and Electronic Systems Magazine}, vol.~39, no.~10, pp. 18--35, 2024.

\bibitem{10908606}
N.~G. Gordon and G.~Falco, ``{An On-Orbit Data Marketplace for Distributed Space Domain Awareness},'' \emph{IEEE Access}, vol.~13, pp. 40\,275--40\,284, 2025.

\bibitem{10535109}
G.~Falco and N.~G. Gordon, ``{A Zero-Trust Satellite Services Marketplace Enabling Space Infrastructure as a Service},'' \emph{IEEE Access}, vol.~12, pp. 71\,066--71\,075, 2024.

\bibitem{9129732}
C.~Fan, S.~Ghaemi, H.~Khazaei, and P.~Musilek, ``{Performance Evaluation of Blockchain Systems: A Systematic Survey},'' \emph{IEEE Access}, vol.~8, pp. 126\,927--126\,950, 2020.

\bibitem{8416434}
F.~M. Benčić and I.~Podnar~Žarko, ``Distributed ledger technology: Blockchain compared to directed acyclic graph,'' in \emph{IEEE 38th International Conference on Distributed Computing Systems (ICDCS)}, 2018, pp. 1569--1570.

\bibitem{9520348}
Z.~Bao, M.~Luo, H.~Wang, K.-K.~R. Choo, and D.~He, ``{Blockchain-Based Secure Communication for Space Information Networks},'' \emph{IEEE Network}, vol.~35, no.~4, pp. 50--57, 2021.

\bibitem{10219188}
D.~Bhuva and S.~Kumar, ``{Securing Space Cognitive Communication with Blockchain},'' in \emph{IEEE Cognitive Communications for Aerospace Applications Workshop (CCAAW)}, 2023, pp. 1--6.

\bibitem{9837853}
Z.~Zhang \emph{et~al.}, ``{A public blockchain consensus mechanism for fault-tolerant distributed computing in LEO satellite communications},'' \emph{China Communications}, vol.~19, no.~7, pp. 110--123, 2022.

\bibitem{REED}
H.~Reed, N.~Dailey, R.~Carden, and D.~Bryson, ``{Blockchain Enabled Space Traffic Awareness (BESTA): Discovery of Anomalous Behavior Supporting Automated Space Traffic Management},'' 2020.

\bibitem{JAIN2025100065}
A.~K. Jain, N.~Gupta, and B.~B. Gupta, ``{A survey on scalable consensus algorithms for blockchain technology},'' \emph{Cyber Security and Applications}, vol.~3, p. 100065, 2025.

\end{thebibliography}

\end{document}